\documentclass[onecolumn,showpacs,superscriptaddress,pre,nofootinbib]{revtex4}

\usepackage{graphicx}
\usepackage{amsmath,amssymb}
\usepackage{color}
\usepackage{amsfonts}

\begin{document}

\newtheorem{theorem}{Theorem}
\newtheorem{definition}{Definition}
\newtheorem{lemma}{Lemma}
\newtheorem{proposition}{Proposition}
\newtheorem{remark}{Remark}
\newtheorem{corollary}{Corollary}
\newtheorem{example}{Example}


\title{L\'evy Transport in Slab Geometry of Inhomogeneous Media}

\author{Alexander Iomin}
\email{iomin@physics.technion.ac.il} \affiliation{Department of Physics, Technion, Haifa 32000, Israel}

\author{Trifce Sandev}
\affiliation{Max Planck Institute for the Physics of Complex Systems,
N\"{o}thnitzer Strasse 38, 01187 Dresden, Germany} \affiliation{Radiation Safety
Directorate, Partizanski odredi 143, P.O. Box 22, 1020 Skopje,
Macedonia}

\date{\today} 

\begin{abstract}
We present a physical example, where a fractional (both in space and time)
Schr\"odinger equation appears only as a formal effective description of
diffusive wave transport in complex inhomogeneous media.
This description is a result of the parabolic equation approximation that corresponds to the paraxial small angle approximation of the fractional Helmholtz equation.
The obtained effective quantum dynamics is fractional in both space and time.
As an example, L\'evy flights in an infinite potential well
are considered numerically. An analytical expression for the effective wave
function of the quantum dynamics is obtained as well.
\end{abstract}

\maketitle

\section{Introduction}

Application of the fractional calculus to quantum processes is a
new and fast developing part of quantum physics, which studies
nonlocal quantum phenomena \cite{kusnezov,laskin1,west,WBG}.
It aims to explore nonlocal effects found for either long-range
interactions or time-dependent processes with many scales
\cite{WBG,bouchaud,klafter}. As it is shown in the seminal papers \cite{laskin1,
west}, the fractional
concept can be introduced in quantum physics by means of the Feynman propagator for
non-relativistic quantum mechanics in complete analogy with the Brownian path integrals \cite{feynman}.
The path integral approach for L\'evy stable processes,
leading to the fractional diffusion equation, can be extended to a
quantum Feynman-L\'evy measure, which leads to natural appearance of the space fractional
Schr\"odinger equation \cite{laskin1,west}.

A way of introduction of a fractional time derivative in quantum mechanics
is not so ``easy''.
It is tempting to introduce it in the quantum
mechanics by analogy with the fractional Fokker-Planck equation (FFPE)
by means of the Wick rotation of time
$t\rightarrow -it/\hbar$ \cite{naber}. However this way is completely vague
from its physical interpretation. First, it violates Stone's theorem on the
one-parameter unitary group\footnote{Stone's theorem on one-parameter unitary groups is a basic theorem of functional analysis that establishes a one-to-one correspondence between self-adjoint operators in the Hilbert space and one-parameter families of unitary operators, which are evolution operators in quantum mechanics. In other words, for the evolution (unitary) operator $\hat{U}(t)$, there is a group property $\hat{U}(t)\hat{U}(s)=\hat{U}(t+s)$.} \cite{stone}.
Another important discrepancy between the Fokker-Planck  and the Schr\"odinger equations is
that the latter is the fundamental quantum-mechanical postulate, while the former is just
an asymptotic limit of a kinetic equation for the Markovian random process.
It is worth noting that, contrary to the space fractional
derivative, the fractional time Schr\"odinger equation (FSE)
describes non-Markovian evolution with a memory effect.
For the Markovian processes the link between the Schr\"odinger equation and the Fokker-Planck equation is established by the path integral
presentation \cite{kac,schulman,chaichian}. Contrary to this for the non-Markovian processes
this path integral relation is broken due to violation of the Stone theorem, therefore the relation between FSE and FFPE is broken as well.

However, these arguments do not discard a mathematical attractivity of the FSE.
As shown in \cite{iom2009,iom2011}, the FSE is an effective way to describe the dynamics of a quantum system interacting with the environment. In this case, fractional time derivative is an effect of the interaction of the quantum system with the environment, where part of the quantum information is lost \cite{iom2011,wo2010}. Therefore, in the paper we consider a physical example of a fractional (both in
space and time) Schr\"{o}dinger equation, which  appears
only as a formal effective description of diffusive wave transport in
complex inhomogeneous media. We show that considering FFPE in the parabolic equation approximation,
which corresponds to the paraxial small angle approximation, one can
obtain the FSE as an effective way
to solve fractional eigenvalue problem in slab geometry, where the paraxial
small angle approximation is naturally applied. The method of parabolic equation approximation
was first applied by Leontovich in study of radio-waves spreading \cite{leontovich}
and later developed in detail by Khokhlov \cite{khohlov} (see also \cite{tappert}).
In modern experimental and theoretical research, it naturally appears in
investigations of a wave propagation in random optical lattices
\cite{moti1,moti2,moti3,moti4}.

The FFPE is one of the most general asymptotic
description of a wave-diffusion transport in inhomogeneous media \cite{WGMN,mainardi1}.
For example, fractional calculus description has been introduced to describe
the wave propagation at anomalous diffusive oscillations in
the Lorentz (Cauchy) pair plasma with dust impurities \cite{turski1},
random walk of optic rays in L\'evy lens \cite{LevyLens},
fractional wave-diffusion in heterogeneous media \cite{mainardi,Meerschaert} as well as
seismic waves \cite{mainardi2,Casasanta}.

This paper is organized as follows. In Section 2 we give a brief introduction to fractional calculus and Mittag-Leffler functions. FFPE in a slab geometry is considered in Section 3. By using paraxial approximation we derive the corresponding FSE as the parabolic equation approximation for the considered process. In Section 4, the obtained model is considered as a
kind of a generalisation of the fractional quantum mechanics. Summary is given in Section 5.


\vspace*{0.5cm}
\setcounter{equation}{0}
\section{Mathematical tools: Fractional calculus briefly}

Fractional derivation was developed as a generalization of integer
order derivatives and is defined as the inverse operation to the
fractional integral. Fractional integration of the order of
$\alpha$ is defined by the operator (see {\em e.g.}
\cite{WBG,klafter,podlubny,oldham,SKM})
\[{}_{a}I_x^{\alpha}f(x)=
\frac{1}{\Gamma(\alpha)}\int_a^xf(y)(x-y)^{\alpha-1}dy\, , \]
where $\alpha>0,~x>a$ and  $\Gamma(z)$ is the Gamma function.
Therefore, the fractional derivative is defined as the inverse
operator to ${}_aI_x^{\alpha}$, namely $
{}_aD_x^{\alpha}f(x)={}_aI_x^{-\alpha}f(x)$ and
${}_aI_x^{\alpha}={}_aD_x^{-\alpha}$. Its explicit form is
\[{}_aD_x^{\alpha}f(x)=
\frac{1}{\Gamma(-\alpha)}\int_a^xf(y)(x-y)^{-1-\alpha}dy\, . \]
For arbitrary $\alpha>0$ this integral diverges, and as a result
of this a regularization procedure is introduced with two
alternative definitions of ${}_aD_x^{\alpha}$. For an integer $n$
defined as $n-1<\alpha<n$, one obtains the Riemann-Liouville
fractional derivative of the form
\begin{equation}\label{mt1a}   %
{}_a^{RL}D_x^{\alpha}f(x)=\frac{d^n}{dx^n}{}_aI_x^{n-\alpha}f(x)\, ,
\end{equation}
and fractional derivative in the Caputo form \cite{mainardi}
\begin{equation}\label{mt1b}  %
{}_a^{C}D_x^{\alpha}f(x)= {}_aI_x^{n-\alpha}f^{(n)}(x)\, , ~~~
f^{(n)}(x)\equiv\frac{d^n}{dx^n}f(x)\, .
\end{equation}  %
There is no constraint on the lower limit $a$. For example, when
$a=0$, one has ${}_0^{RL}D_x^{\alpha}x^{\beta}=\frac{x^{\beta-\alpha}
\Gamma(\beta+1)}{\Gamma(\beta+1-\alpha)}$ and
$${}_0^{C}D_x^{\alpha}f(x)=
{}_0^{RL}D_x^{\alpha}f(x)-\sum_{k=0}^{n-1}f^{(k)}(0^+)
\frac{x^{k-\alpha}}{\Gamma(k-\alpha+1)}\, ,$$ and
${}_a^{C}D_x^{\alpha}[1]=0$, while
${}_0^{RL}D_x^{\alpha}[1]=x^{-\alpha}/\Gamma(1-\alpha)$. When
$a=-\infty$, the resulting Weyl derivative is
\begin{equation}\label{weyl1}
{}_{-\infty}\mathcal{W}_x^{\alpha}\equiv{}_{-\infty}^{W}D_x^{\alpha}=
{}_{-\infty}^{RL}D_x^{\alpha}= {}_{-\infty}^{C}D_x^{\alpha}\, .
\end{equation}
One also has ${}_{-\infty}^{W}D_x^{\alpha}e^x=e^x$ This property is
convenient for the Fourier transform
\begin{equation}\label{top_0a}
\mathcal{F}\left[{}_{-\infty}\mathcal{W}_x^{\alpha}f(x)\right]=(ik)^{\alpha}\bar{f}(k)\, ,  %
\end{equation}  %
where $\mathcal{F}[f(x)]=\bar{f}(k)$. This fractional derivation with the
fixed lower limit is also called the left fractional derivative.
One can introduce the right fractional derivative, where
the upper limit $a$ is fixed and $a>x$. For example, the right
Weyl derivative is
\begin{equation}\label{top_0b}
{}_x\mathcal{W}_{\infty}^{\alpha}f(z)=\frac{1}{\Gamma(-\alpha)}
\int_x^{\infty}\frac{f(y)dy}{(y-x)^{1+\alpha}}\, .
\end{equation}  %
The Laplace transform of the Caputo fractional derivative yields
\begin{equation}\label{mt2}
 \mathcal{L}[{}_0^{C}D_x^{\alpha}f(x)]=
s^{\alpha}\tilde{f}(s)-\sum_{k=0}^{n-1}f^{(k)}(0^+)s^{\alpha-1-k}
\, ,
\end{equation}
where $\mathcal{L}[f(x)]=\tilde{f}(s)$, that is convenient for the
present analysis, where the initial conditions are supposed in
terms of integer derivatives. We also use here a convolution rule
for $0<\alpha<1$
\begin{equation}\label{mt3}
\mathcal{L}[{}_0I_x^{\alpha}f(x)]=s^{-\alpha}\tilde{f}(s)\, .
\end{equation}
The fractional derivative from an exponential function can be
simply calculated as well by virtue of the Mittag--Leffler
function (see {\em e.g.}, \cite{podlubny,bateman}):
\begin{equation}\label{mt4}   %
E_{\gamma,\delta}(z)=\sum_{k=0}^{\infty}
\frac{z^k}{\Gamma(\gamma k+\delta)} \, ,
\end{equation}
where $\gamma,\delta>0$ and for $\delta=1$ one also obtains the Mittag--Leffler
function in one parameter $E_{\gamma,1}(z)=E_{\gamma}(z)$ \cite{podlubny}.
Therefore, we have the following expression
\begin{equation}\label{mt5}   %
{}_0^{RL}D_x^{\alpha}e^{\lambda x}=x^{-\alpha}E_{1,1-\alpha}(\lambda x)\, .
\end{equation}
while for the Caputo fractional derivative the Mittag-Leffler function $E_{\gamma}(z^{\gamma})$
is the eigenfunction
\begin{equation}\label{mt6}   %
{}_0^{C}D_z^{\gamma}E_{\gamma}(\lambda z^{\gamma})=\lambda E_{\gamma}(\lambda z^{\gamma})\, .
\end{equation}
The Laplace transform of the Mittag-Leffler function (\ref{mt4}) is given by \cite{podlubny}
\begin{equation}\label{ml2 laplace}   %
\mathcal{L}\left[z^{\delta-1}E_{\gamma,\delta}(\pm \omega z^{\gamma})\right](s)=\frac{s^{\gamma-\delta}}{s^{\gamma}\mp\omega}, \quad \Re(s)>|\omega|^{1/\gamma}.
\end{equation}
For the Mittag-Leffler function (\ref{mt4}) the following formula holds \cite{bateman}
\begin{equation}\label{ML two asymptotic}
E_{\gamma,\delta}(-z)=-\sum_{n=1}^{\infty}\frac{(-z)^{-n}}{\Gamma(\delta-\beta n)}, \quad |z|>1,
\end{equation}
from where it follows the following asymptotic behavior
\begin{equation}\label{ML two asymptotic2}
E_{\gamma,\delta}(-z)=\frac{z^{-1}}{\Gamma(\delta-\beta)}, \quad |z|\rightarrow\infty.
\end{equation}


\vspace*{0.5cm}
\setcounter{equation}{0}
\section{FFPE in slab geometry: parabolic equation approximation}

We begin our analysis by considering a general form of the FFPE
for the 2D slab geometry described by
dimensionless $(r,z)$ variables, where $z\in (0,\,\infty)$ and $r\in [-L,\,L]$. Therefore
the probability distribution function $P(r,z,t)$ to find a particle at the point $(r,z)$ at time $t$ is governed by the FFPE,
which consists of both the Caputo and the Riemann-Liouville fractional derivatives.
In what follows, the low limit of the Caputo derivatives is $a=0$,
therefore, we use the following notation
\begin{equation}\label{ffpe_mta}
\partial_x^{\alpha}\equiv {}_0^{C}D_x^{\alpha}\, , ~~~x=z,\, t\, .
\end{equation}
These fractional derivatives are used for the dimensionless time $t\in(0,\, \infty)$ and the longitudinal direction $z(0,\, \infty)$. In this case the Laplace transform (\ref{mt2}) is
determined explicitly by the initial condition $P(r,z,t=0)=P_0(r,z)$ and its evolution
in time at $z=0$.
For the orthogonal direction $r\in[-L,\, L]$, the following notation is used
\begin{equation}\label{ffpe_mtb}
\mathcal{D}_{|x|}^{\alpha}\equiv {}_{-L}^{RL}D_x^{\alpha}+
{}_{x}^{RL}D_{L}^{\alpha}\, , ~~ x=r\, .
\end{equation}
The FFPE reads
\begin{equation}\label{ffpe_1}
\partial_t^{\alpha}P=\mathcal{K}_{\beta}\partial_z^{\beta}P+
\mathcal{K}_{\beta}\mathcal{D}_{|r|}^{\beta}P\, ,
\end{equation}
where $\mathcal{K}_{\beta}$ is a diffusivity of the media, while $0<\alpha<1$ and $1<\beta<2$.

The FFPE (\ref{ffpe_1}) is a  general form of space-time fractional equations.
It describes both waves and relaxation processes in a variety of applications like
diffusion-wave phenomena in inhomogeneous media \cite{mainardi1,turski1,LevyLens,mainardi,Meerschaert,mainardi2}.
It should be stressed that fractional space derivatives describe L\'evy flights \cite{klafter}.
In particular, we specify here optical ray dynamics
in L\'evy glasses \cite{LevyLens}, where L\'evy flights can be described by Eq. (\ref{ffpe_1})\footnote{L\'evy glasses are specially prepared optical material in which
the L\'evy flights are controlled by the power law distribution of the step-length of a
free ray dynamics, which can be specially chosen in the power law form $\sim 1/l^{\beta+1}$.}.
Another interesting phenomenon, which is described by Eq. (\ref{ffpe_1}), is superdiffusion of
ultra-cold atoms in an optical lattice  \cite{nirD}\footnote{There are L\'evy walks, and the theoretical explanation of this fact, presented within the standard semiclassical treatment of Sisyphus cooling \cite{zoller,eli1}, is based on a
study of the microscopic characteristics of the atomic motion in optical lattices and recoil distributions resulting in macroscopic L\'evy walks in space, such that the L\'evy distribution
of the flights depends on the lattice potential depth \cite{zoller}.
The flight times and velocities of atoms are coupled, and these relations, established in asymptotically logarithmic potential, have been studied for
different regimes of the atomic dynamics, \cite{eli1,eli2}, so the cold atom problem is a variant of the L\'evy walks.}

Therefore, our main concern is now the Helmholtz fractional
equation, which relates to the L\'evy process.  The next step
of the analysis is the separation of variables (see e.g. \cite{klafter})
according the following separation ansatz
\begin{equation}\label{ffpe_2}
P(r,z,t)=\int_{-i\infty+\sigma}^{+i\infty+\sigma}E_{\alpha}(\omega t^{\alpha}) W(r,z,\omega)d\omega\, ,
\end{equation}
where $E_{\alpha}(\omega t^{\alpha})$ is the Mittag-Leffler function (\ref{mt4}), while
$W(r,z,\omega)$ is determined from the fractional Helmholtz equation
\begin{equation}\label{ffpe_3}
\partial_z^{\beta}W+\mathcal{D}_{|r|}^{\beta}W+\omega_{\beta}W=0\, ,
\end{equation}
where $\omega_{\beta}=\frac{\omega}{\mathcal{K}_{\beta}}$.
When the height $L$ is less than the L\'evy flight lengths in the longitudinal direction, the transport is of a small grazing angle with respect to the longitudinal direction.
We are looking for the solution $W(z,r,\omega)$ in the form
\begin{equation}\label{ffpe_4}
W(z,r.\omega)=e^{ikz}u(z,r,\omega),
\end{equation}
which after substitution in Eq. (\ref{ffpe_3}) yields the following integration
\begin{equation}\label{ffpe_5}
\partial_{z}^{\beta}W(z,r,\omega)=\frac{1}{\Gamma(2-\beta)}
\int_0^z(z-z')^{2-\beta-1}\frac{d^2}{dz'^2}\left[e^{ikz'}u(z',r,\omega)\right]dz\, .
\end{equation}
Note that the ``initial'' conditions at $z=0$ for both $u$ and $W$ are the same:
$u(z=0)=W(z=0)$.

Now the parabolic equation in the paraxial approximation can be obtained.
Taking into account that $u(z,r,\omega)$ is a slowly-varying function of $z$, such that
\begin{equation}\label{ffpe_5_6}
\Big|\frac{\partial^2 u}{\partial z^2}\Big|\ll\Big|2k\frac{\partial u}{\partial z}\Big|\, ,
\end{equation}
one obtains
\begin{equation}\label{ffpe_6}
 \frac{d^2}{dz^2}\left[e^{ikz}u(z,r,\omega)\right]\approx
 2ike^{ikz}\frac{d}{dz}u(z,r,\omega)\, .
 \end{equation}
After substitution of this approximation in (\ref{ffpe_5}), one obtains
from Eqs.~(\ref{ffpe_3}), (\ref{ffpe_4}), and (\ref{ffpe_5})
\begin{equation}\label{ffpe_7}
2ik\, {}_0I_z^{2-\beta}\left[e^{ikz}\partial_zu(z,r,\omega)\right]+
\mathcal{D}_{|r|}^{\beta}u(z,r,\omega)e^{ikz}+\omega_{\beta}u(z,r,\omega)e^{ikz}=0\, .
\end{equation}
To get rid of the exponential $e^{ikz}$ in Eq.~(\ref{ffpe_7}), we insert it
inside the derivative $e^{ikz}\partial_zu(z,r,\omega)=\partial_z[e^{ikz}u(z,r,\omega)]+{\rm O}(k)$. The term ${\rm O}(k)=ik e^{ikz}u(z,r,\omega)$ can be neglected
in Eq.~(\ref{ffpe_7}) since it is of the order of ${\rm O}(k^2)$, which
is neglected in the paraxial approximation with $k\ll 1$. Note also that
${}_0I_z^{2-\beta}\frac{d}{dz}f(z)=\partial_z^{\gamma}f(z)$, where $\gamma=\beta-1$ and $0<\gamma<1$.
The Laplace transform can be performed: $\mathcal{L}[e^{ikz}u(z)]=\tilde{u}(s-ik)$. Therefore, one obtains from Eq.~(\ref{ffpe_7})
\begin{equation}\label{ffpe_8}
2ik[s^{\gamma}\tilde{u}(s-ik)-s^{\gamma-1}u(z=0)]+\mathcal{D}_{|r|}^{\beta}\tilde{u}(s-ik)+
\omega_{\beta}\tilde{u}(s-ik)\, .
\end{equation}
Performing the shift $s-ik\rightarrow s$ and neglecting again the terms of the order of ${\rm o}(k)$ in Eq.~(\ref{ffpe_8}), and then performing the Laplace inversion, one obtains the
Helmholtz equation in
the form of the effective fractional Schr\"odinger equation (FSE),
where the $z$ coordinates play a role of an effective time
\begin{equation}\label{ffpe_9}
2ik\partial_z^{\gamma}u+\mathcal{D}_{|r|}^{\beta}u+\omega_{\beta}u=0\, .
\end{equation}
The ``initial'' condition at $z=0$ corresponds to the boundary condition for
the initial problem in Eq.~(\ref{ffpe_1}) for the distribution $P(r,z,t)$.
One supposes that there is a source of the signal at $z=0$, therefore we have
the initial condition $u(z=0,r)=P_0(r,0)$. The boundary conditions
at $r=\pm L$ are $u(r=\pm L,z)=0$.


\vspace*{0.5cm}
\setcounter{equation}{0}
\section{FSE of a ``free'' particle in the infinite well potential}

In general case, the diffusion coefficient $\mathcal{K}_{\beta}$ is not a constant value,
therefore $\omega_{\beta}$ plays a role of potential. However, even for $\mathcal{K}_{\beta}$
being a constant, a quantum particle is not free, it moves in the infinite potential well. Under these conditions, Eq.~(\ref{ffpe_9}) is a kind of a generalisation of the fractional quantum mechanics, considered in Refs.~\cite{saxena1,saxena2}. One should bear in mind that the obtained
FSE (\ref{ffpe_9}) is an \textit{effective} quantum mechanics, which is the result of the parabolic equation   approximation of the initial Helmholtz equation (\ref{ffpe_3}).
In this sense, use of the quantum mechanical terminology, like wave function,
is formal, since the obtained equation leads to quantum mechanical paradigms.

Here we consider the case $\mathcal{K}_{\beta}={\rm const}$. Then $\omega_{\beta}$ can be omitted
from the fractional Hamiltonian $\hat{H}=\mathcal{D}_{|r|}^{\beta}$, and Eq.~(\ref{ffpe_9}) splits into two eigenvalue equations
\begin{equation}\label{fse_1a}
2ik\partial_z^{\gamma}Z(z) + \omega_{\beta}Z(z)=-eZ(z)\, ,
\end{equation}
\begin{equation}\label{fse_1b}
\mathcal{D}_{|r|}^{\beta}R(r)=eR(r)\, ,
\end{equation}
where $u(z,r)=Z(z)R(r)$ and $e$ is the eigenspectrum of $\hat{H}$, which corresponds to a fractal dynamics of a particle (namely, L\'evy flights) in a box \cite{iomin_2015}. The initial condition
for $Z(z)$ is $Z(0)=Z_0$, while the boundary condition for $R(r)$ are $R(r=\pm L)=0$.

\subsection{Solution of the initial value problem}

The solution of Eq.~(\ref{fse_1a}) can be obtained by employing the Laplace transform method. Thus, from relation (\ref{mt2}) one finds
\begin{equation}\label{Eq23 Laplace1}
s^{\gamma}\tilde{Z}(s)-s^{\gamma-1}Z(0)+\frac{\omega_{\beta}+e}{2\imath k}\tilde{Z}(s)=0\, ,
\end{equation}
from where it follows
\begin{equation}\label{Eq23 Laplace2}
\tilde{Z}(s)=\frac{s^{\gamma-1}}{s^{\gamma}+\frac{\omega_{\beta}+e}{2\imath k}}Z(0)\, .
\end{equation}
After the Laplace inversion, according relation (\ref{ml2 laplace}), the solution
of Eq.~(\ref{fse_1a}) reads
\begin{equation}\label{Eq23 solutionZ}
Z(z)=Z(0)E_{\gamma}\left(-\frac{\omega_{\beta}+e}{2\imath k}z^{\gamma}\right)=Z(0)E_{\gamma}\left(\imath\frac{\omega_{\beta}+e}{2k}z^{\gamma}\right)\, .
\end{equation}
The solution can be represented in a form $Z(z)=\mathcal{R}(z)+\imath\mathcal{I}(z)$, i.e.,
\begin{align}\label{Eq23 solutionZfinal}
Z(z)=Z(0)&\left[E_{2\gamma}
\left(-\left(\frac{\omega_{\beta}+e}{2k}\right)^{2}z^{2\gamma}\right)\right.\nonumber\\
&\left.+\imath\frac{\omega_{\beta}+e}{2k}z^{\gamma}E_{2\gamma,\gamma+1}
\left(-\left(\frac{\omega_{\beta}+e}{2k}\right)^{2}z^{2\gamma}\right)\right]\, .
\end{align}
From (\ref{Eq23 solutionZfinal}) it follows
\begin{align}\label{Eq23 solutionZfinal2}
|Z(z)|^{2}=Z^{2}(0)&\left[\left(E_{2\gamma}
\left(-\left(\frac{\omega_{\beta}+e}{2k}\right)^{2}z^{2\gamma}\right)\right)^{2}\right.\nonumber\\
&\left.+\left(\frac{\omega_{\beta}+e}{2k}\right)^{2} z^{2\gamma}\left(E_{2\gamma,\gamma+1}
\left(-\left(\frac{\omega_{\beta}+e}{2k}\right)^{2}z^{2\gamma}\right)\right)^{2}\right]\, .
\end{align}

Taking into account definition (\ref{mt4})
of the Mittag-Leffler function, one obtains the initial behavior for $z\rightarrow 0$
\begin{align}\label{Eq23 solutionZfinal z0}
|Z(z)|^{2}&\simeq Z^{2}(0)
\left[\left(1-\frac{\left(\frac{\omega_{\beta}+e}{2k}\right)^{2}z^{2\gamma}}
{\Gamma\left(2\gamma+1\right)}\right)^{2}\right.\nonumber\\
&\left.+\left(\frac{\omega_{\beta}+e}{2k}\right)^{2}z^{2\gamma}\left(\frac{1}{\Gamma(\gamma+1)}
-\frac{\left(\frac{\omega_{\beta}+e}{2k}\right)^{2}z^{2\gamma}}
{\Gamma\left(3\gamma+1\right)}\right)^{2}\right]\nonumber\\
&\simeq Z^{2}(0)
\left[1-2\frac{\left(\frac{\omega_{\beta}+e}{2k}\right)^{2}z^{2\gamma}}
{\Gamma\left(2\gamma+1\right)}+\frac{\left(\frac{\omega_{\beta}+e}{2k}\right)^{2}z^{2\gamma}}
{\Gamma^{2}(\gamma+1)}\right]\, .
\end{align}
In the limit $z\rightarrow\infty$, we find by employing relation (\ref{ML two asymptotic2})  that the norm is not conserved, i.e., it has a power-law decay of the form
\begin{equation}\label{Eq23 solutionZfinal z large}
|Z(z)|^{2}\simeq Z^{2}(0)\frac{\left(\frac{\omega_{\beta}+e}{2k}\right)^{-2}
z^{-2\gamma}}{\Gamma^{2}\left(1-\gamma\right)}\, .
\end{equation}

\subsection{Solution of the boundary problem}

As it is shown in Ref.~\cite{iomin_2015}, Eq.~(\ref{fse_1b}) describes quantum L\'evy flights in the infinite potential well, such that ``free'' L\'evy motion takes place inside the segment $[-L,L]$ only. In other words, the fractional operator of the kinetic energy is defined on this segment only, and any discussion about the wave function outside of the segment $[-L,L]$ is irrelevant. This is a typical example of quantum mechanics with a topological constraint \cite{schulman,chaichian,iomin_2015}. This eigenvalue problem has been solved analytically \cite{iomin_2015}. In the present study of the slab dynamics, we use the odd
eigenfunction\footnote{Note that this basis is complete.}, obtained numerically and shown
in Fig.~\ref{fig1}, and their analytical inferring is presented in Appendix A.
The antisymmetric (odd) eigenfunction reads
\begin{equation}\label{eig3}  %
R_e^{\rm o}(r)=\Psi_m^{\rm odd}(r)=\frac{1}{\sqrt{L}}\sin\frac{m\pi r}{L}\, , ~~~~m=1,2, \dots,
\end{equation}
which satisfies the boundary condition $\Psi_m^{\rm odd}(r=\pm L)=0$ and corresponds to the eigenvalue $e_m^{\rm o}=\left(\frac{m\pi}{L}\right)^{\beta}$. Therefore, the general solution of Eq. (\ref{ffpe_9}) reads
\begin{equation}\label{solution_ffpe}
u(z,r)=\sum_{m=1}^{\infty}A_m
E_{\gamma}\left(\imath\frac{\omega_{\beta}+e_m}{2k}z^{\gamma}\right)
\sin\frac{m\pi r}{L}\, ,
\end{equation}
where coefficients of the expansion $A_m$ are determined from the initial condition
$u(z=0,r)=\sum_{m=1}^{\infty}A_m\sin\frac{m\pi r}{L}$.

We suggest a graphical way of the solution of Eq.~(\ref{fse_1b}).
The numerical procedure of calculation of the fractional derivative $\mathcal{D}_{|r|}^{\beta}$
is based on the fast Fourier transform (FFT). We found numerically that this procedure
works perfectly for the odd eigenfunctions
of Eq.~(\ref{eig3}) and yields the correct spectrum\footnote{The FFT
procedure does not work for the even eigenfunctions. The symmetric (even)
eigenfunctions, obtained in Ref.~\cite{iomin_2015}, are
$R_e^{\rm e}(r)=\Psi_{2m+1}^{\rm even}(r)
=\frac{1}{\sqrt{L}}\cos\left[\frac{(2m+1)\pi}{2L}r\right]$, and their spectrum is
$e_m^{\rm e}=\left[\frac{(2m+1)\pi}{2L}\right]^{\beta}$.}.
Graphical solutions $R_{m}^{o}(r)$ for the first three states obtained by the FFT procedure, are shown in Fig.~\ref{fig1}. We also found a class of degenerate functions
 with the spectrum $e_m=\left(\frac{m\pi}{L}\right)^{\beta}$, where for each $m$ there are two normalized functions
 $\Psi_m^{c}(r)=\frac{1}{\sqrt{L}}\cos\left(\frac{m\pi (2r+L)}{2L}\right)$ and $\Psi_m^{s}(r)=\frac{1}{\sqrt{L}}\sin\left(\frac{m\pi(2r+L)}{2L}\right)$. However, the boundary conditions lift this degeneracy, such that for $m=2n$,  one obtains $R_m(r)=\Psi_m^{s}(r)$,
 while for $m=2n+1$ one obtains $R_m(r)=\Psi_m^{c}(r)$ (see Fig.~\ref{fig2}).

\begin{figure}[H]
\centering
(a) \includegraphics[width=9.0cm]{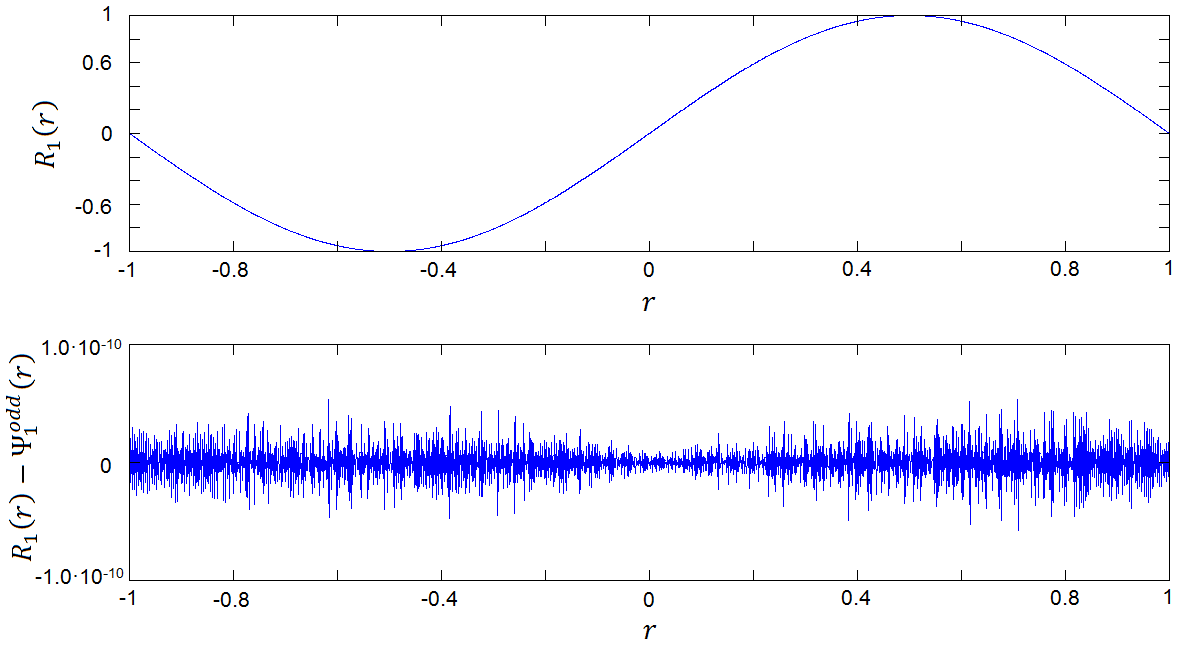}\\
(b) \includegraphics[width=9.0cm]{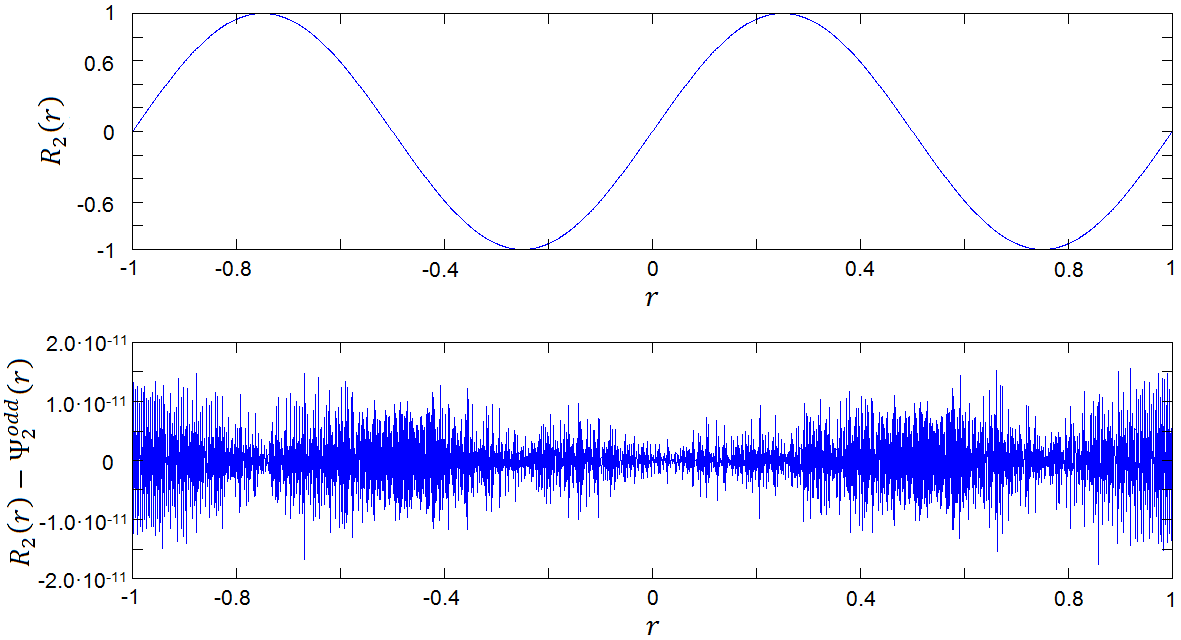}\\
(c) \includegraphics[width=9.0cm]{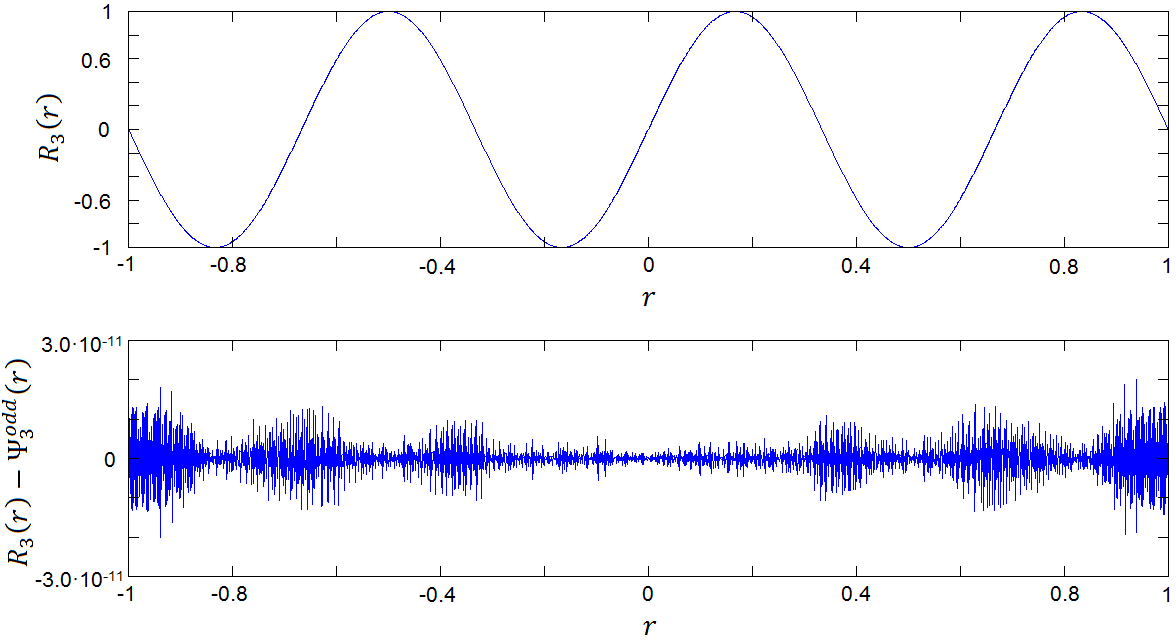}\\

 \caption{Graphical presentation of the first three odd eigenstates $R_{m}^{o}(r)= \sin(m\pi r/L)\, , ~~m=1,2,3$, for $\beta=1.8$ and $L=1$,
 obtained by FFT. From the plots of $R_{m}^{o}(r)-\Psi_{m}^{odd}(r)\, , ~m=1,2,3$,
 one can conclude that there is a very good agreement with the analytical
 eigenfunctions (\ref{eig3}).}
\label{fig1}
\end{figure}

\begin{figure}[H]
\centering
(a) \includegraphics[width=9.0cm]{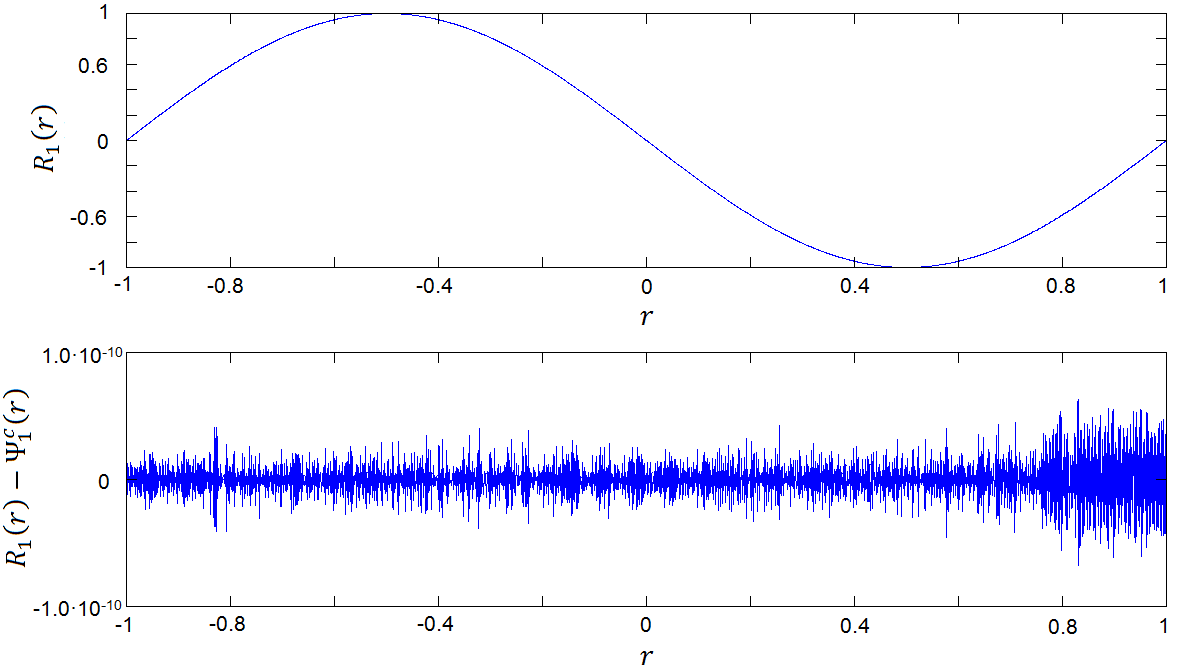}\\
(b) \includegraphics[width=9.0cm]{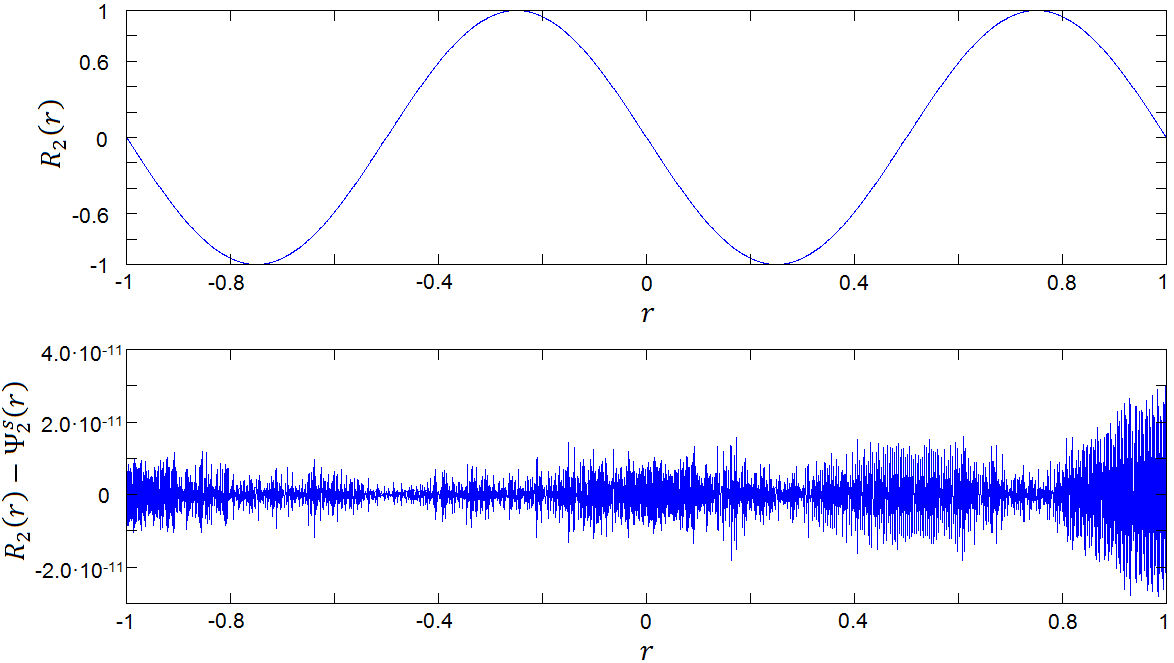}\\
(c) \includegraphics[width=9.0cm]{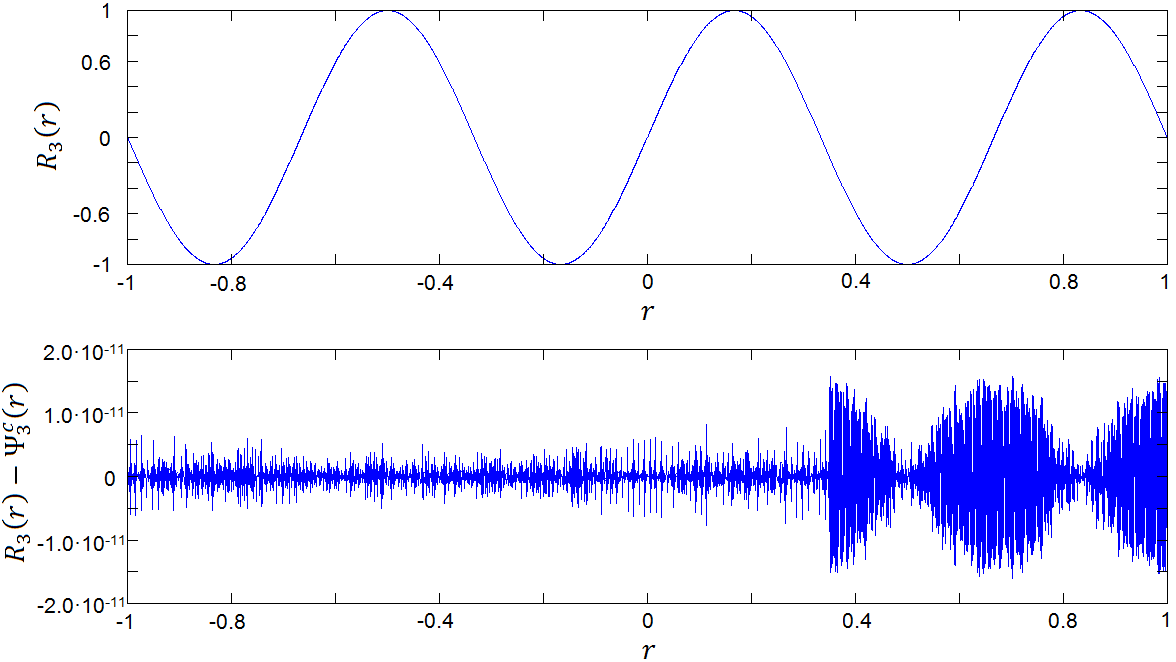}\\
(d) \includegraphics[width=9.0cm]{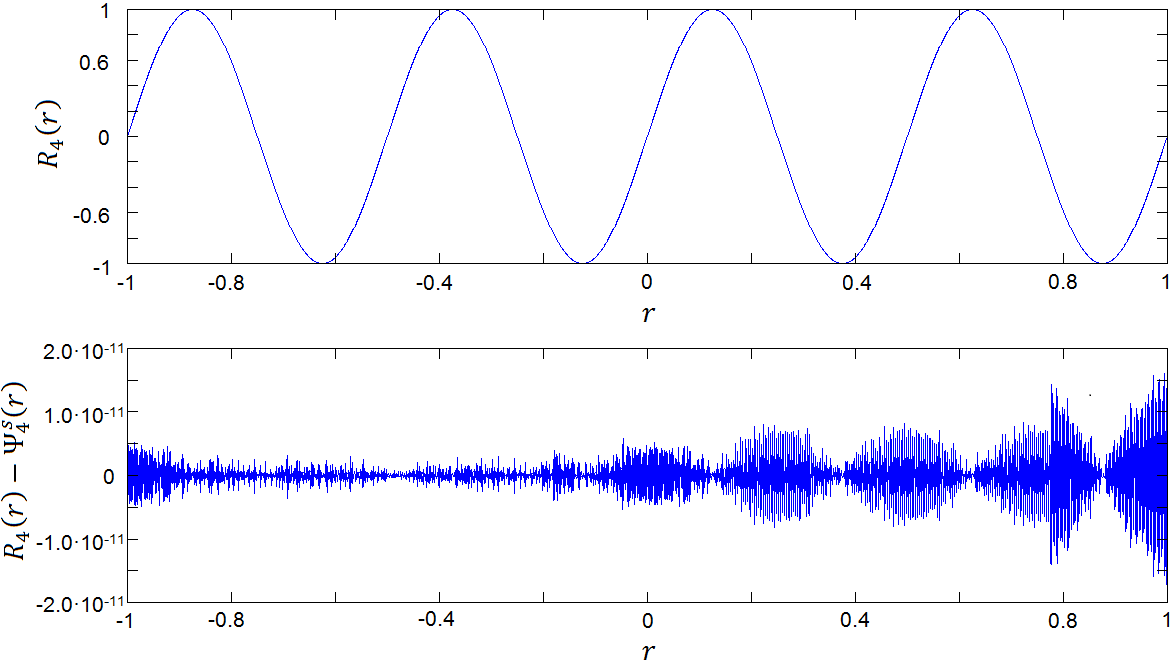}\\

 \caption{Graphical presentation of first fourth states for
 $\beta=1.8$ and $L=1$ for a class of degenerate functions
with the spectrum $e_m=(\pi m/L)^{\beta}$, which corresponds to the odd eigenvalues.
Here for each $m$, there are two normalized functions
$\Psi_m^{c}(r)=\frac{1}{\sqrt{L}}\cos[m\pi(2r+L)/(2L)]$ and
$\Psi_m^{s}(r)=\sin[m\pi(2r+L)/(2L)]$. However, the boundary conditions lift
this degeneracy, such that for $m=2n$,  one obtains
$R_m(r)=\Psi_m^{s}(r)$, while for $m=2n+1$ one obtains $R_m(r)=\Psi_m^{c}(r)$.}
\label{fig2}
\end{figure}


\setcounter{equation}{0}
\section{Summary}

We concerned with fractional wave-diffusion processes, which take place in
inhomogeneous composite media. We described them in the framework of a
fractional Fokker-Planck equation (FFPE). A general form of the real-time-space
FFPE (\ref{ffpe_1}) describes both waves and relaxation processes in a variety of applications of
diffusion-wave phenomena in inhomogeneous media \cite{mainardi1,turski1,LevyLens,mainardi,Meerschaert,mainardi2,nirD}.
The suggested scenario of the analysis is as follows.
We started from a general form of the real-time-space FFPE, which however permits the variable separation. For the two-dimensional slab geometry, we developed a parabolic
equation approximation which reduces the FFPE (\ref{ffpe_1}) to the space-time
fractional Schr\"odinger equation (FSE). Then, since the obtained equation relates to a variety
of quantum paradigms, one can employ the quantum mechanical terminology,
noting that this use is formal only.
The FSE (\ref{ffpe_9}) governs a
L\'{e}vy transport in the slab geometry of inhomogeneous composite media, and
in particular, it describes experimental realization of superdiffusive ray
dynamics in L\'evy glasses \cite{LevyLens} and superdiffusion of ultra-cold atoms
in an optical lattice \cite{nirD}, where for these  different processes, the
L\'evy walks are described  by the same power law distribution $1/r^{1+\beta}$.
For the space-time FSE (\ref{ffpe_9}), there is an effective time,
which relates to the Caputo time fractional derivative and corresponds to the longitudinal direction of the real space of the slab geometry.

In the effective quantum dynamics according to the FSE,
the Caputo fractional time derivative is responsible for the relaxation temporal behavior
due to non-unitary evolution according the Mittag-Lefler functions
in the solution (\ref{Eq23 solutionZ}). The transverse coordinates are responsible
for the L\'evy process by means of the Riesz fractional derivative in the infinite potential well, which is determined by the boundary conditions. This case of a ``free L\'evy'' particle in the infinite potential well is described by Eq.~(\ref{fse_1b}) and it was analyzed numerically, and a correspondence of the numerical results with those obtained analytically in
Ref.~\cite{iomin_2015} is observed.

Regarding Eq.~(\ref{fse_1b}), few words are in order. As it is shown in \cite{iomin_2015},
the space fractional operator exists on the segment $[-L,L]$ only, and it corresponds to
the quantum mechanics with a topological constraint.
However, in addition to the introduction of fractional quantum mechanics by means of the
L\'evy-Feynman measure \cite{laskin1}, here the fractional quantum mechanics appears
also as the result of the parabolic equation approximation in both space and time
for the FFPE, which is specified by the finite boundary conditions.
The introduction of the fractional operator of the kinetic energy in the
infinite well potential on the scale $[-L,L]$ is natural, and any discussion about the wave
function outside of this scale is irrelevant. As the result, the wave function
can be periodically extended on the entire axis $(-\infty\,,\infty)$ that leads to the
periodic solution, which is proven here numerically. We obtained graphical presentation of
the odd eigenfunctions $R_{m}^{o}(r)=\frac{1}{\sqrt{L}}\sin\frac{m\pi r}{L}$ by
using the fast Fourier transform (FFT) as a standard intrinsic function
of the MATLAB package. It is worth noting that although this algorithm is
not valid for the
even eigenfunction\footnote{This problem does not relate to the fractional calculus. It
does not work even for $\beta=2$.},
the obtained correspondence between analytical result and numerical,
graphical calculations proves that the eigenfunctions of the L\'evy flights
in a box are periodic functions (see discussion in Ref.~\cite{iomin_2015}).

\vspace*{0.5cm}

\section*{Acknowledgements}
This research was supported by the Israel Science Foundation (ISF-1028).

\appendix

\vspace*{0.5cm}
\setcounter{equation}{0}
\section{Eigenvalue problem}\label{sec:app_A}

\def\theequation{A. \arabic{equation}}
\setcounter{equation}{0}
In this appendix, we present a result obtained in Ref.~\cite{iomin_2015}, where the eigenvalue problem of Eq. (\ref{fse_1b}) has been considered for the  fractional Laplace operator and the antisymmetric (odd) eigenfunctions $R_e^{\rm o}$ was found in the form
\begin{equation}\label{App_1}  %
R_e^{\rm o}(x)=\Psi_m^{\rm odd}(x)
=\frac{1}{\sqrt{L}}\sin\frac{m\pi x}{L}\, , ~~~~m=1,2,
\dots,
\end{equation} %
which satisfies the boundary condition $\Psi_m^{\rm odd}(x=\pm L)=0$ and $x=r=\pm L$.
We rewrite the fractional Laplace operator in the form
(\ref{ffpe_mtb}), which is convenient in the analysis,
\begin{equation}\label{App_2} %
\hat{H}\Psi_m^{\rm odd}(x)=
(i\partial_x)\int_{-L}^{L}\left[\frac{1}{2\pi}\int_{-\infty}^{\infty}
|k|^{\beta-2}ke^{-ik(x-y)}dk\right]\Psi_m^{\rm odd}(y)dy\,.
\end{equation}%
Let us first perform integration over $y$. This yields
\begin{equation}\label{App_3}  %
\frac{1}{2i}\int_{-L}^{L}\left[e^{iky+iz}-e^{iky-iz}\right]
=(-1)^m\frac{2\pi m}{iL}\frac{\sin(kL)}{(k+z)(k-z)}\, ,
\end{equation} %
where $z=\pi my/L$. The next step is integration over $k$.  From
Eqs. (\ref{App_2}) and (\ref{App_3}), we have integrals
\begin{equation}\label{App_4}  %
\frac{1}{2\pi} \int_{-\infty}^{\infty}
\frac{k|k|^{\alpha-2}}{(k+z)(k-z)}\left[e^{ik(L-x)}-e^{-i(L+x)}\right]
\equiv I^{(+)}-I^{(-)},
\end{equation}  %
where sign $(+)$ corresponds to the analytical continuation in the
upper half plain, while $(-)$ corresponds to the analytical
continuation in the lower half plain. Using the Residue theorem,
one obtains that the integration yields
\begin{equation}\label{App_5} %
I^{(+)}-I^{(-)}=i(-1)^m\left(\frac{m\pi}{L}\right)^{\beta-2}
\cos\left(\frac{m\pi}{L}x\right).
\end{equation}  %
Acting on this result by the rest part of the operator, which
reads
$(-1)^m\left(\frac{m\pi}{L}\right)(i\partial_x)$,
one obtains
\begin{equation}\label{App_6} %
\hat{H}\Psi_m^{\rm odd}(x)=
\left(\frac{m\pi}{L}\right)^{\beta}
\Psi_m^{\rm odd}(x)\equiv e_m^{\rm odd}\Psi_m^{\rm odd}(x). \end{equation}  %



\end{document}